\documentclass[twocolumn]{aastex63}
\usepackage{hyperref}
\usepackage{bm}
\usepackage{multirow}
\usepackage{gensymb}
\usepackage{enumitem}
\usepackage{tabularx}
\usepackage[flushleft]{threeparttable}
\usepackage[]{color}

\usepackage{natbib}
\usepackage{float}

\usepackage{amsfonts,amsmath,amssymb}
\usepackage{graphicx}
\usepackage{color}
\usepackage{xcolor}
\usepackage{natbib}
\usepackage{url}

\newcommand{\E}[1]{\left\langle#1\right\rangle}

\def\nux{{\nu_{\rm h}}}

\newcommand{\pdif}[2]{\ensuremath{ \frac{\partial #1}{\partial #2}}}

\def\<{\,\langle\langle}
\def\>{\,\rangle\rangle}

\defcitealias{Dannen20}{D20}
\defcitealias{Waters21}{WPD21}
\defcitealias{Begelman83}{BMS83}
\def\CIVdbl{{\rm C~}\kern 0.1em{\sc iv}~$\lambda\lambda 1548, 1550$} 
\def\OVIIIi{\hbox{{\rm O}\kern 0.1em{\sc viii}}}
\def\SiXIVi{\hbox{{\rm Si}\kern 0.1em{\sc viii}}}
\def\OVIII{\hbox{{\rm O}\kern 0.1em{\sc viii}~{\rm Ly}$\alpha$}}
\def\SiXIV{\hbox{{\rm Si}\kern 0.1em{\sc viii}~{\rm Ly}$\beta$}}
\def\FeXXV{\hbox{{\rm Fe}\kern 0.1em{\sc xxv}}}
\def\FeXXVI{\hbox{{\rm Fe}\kern 0.1em{\sc xxvi}}}
\def\FeXXVK{\hbox{{\rm Fe}\kern 0.1em{\sc xxv}~{\rm K}$\alpha$}}
\def\FeXXVIK{\hbox{{\rm Fe}\kern 0.1em{\sc xxvi}~{\rm K}$\alpha$}}

\newcommand{\beq}{\begin{equation}}
\newcommand{\seq}{\end{equation}}

\newcommand{\gv}[1]{\ensuremath{\mbox{\boldmath$ #1 $}}} 
\newcommand{\pd}[2]{\frac{\partial #1}{\partial #2}}
\newcommand{\pdtext}[2]{\partial #1/\partial #2}
\newcommand{\f}{\frac}  

\usepackage{letltxmacro,amsmath}
\LetLtxMacro{\originaleqref}{\eqref}
\renewcommand{\eqref}{Eq.~\originaleqref}

\submitjournal{ApJL}
\shorttitle{ti in rhd}
\shortauthors{Proga et al.}
\begin{document}
\title{Thermal instability in radiation hydrodynamics: \\
instability mechanisms, position-dependent S-curves, and attenuation curves}

\correspondingauthor{Daniel Proga}
\email{daniel.proga@unlv.edu}
\author[0000-0002-6336-5125]{Daniel Proga}
\affiliation{Department of Physics \& Astronomy \\
University of Nevada, Las Vegas \\
4505 S. Maryland Pkwy \\
Las Vegas, NV, 89154-4002, USA}
\author[0000-0002-5205-9472]{Tim Waters}
\affiliation{Department of Physics \& Astronomy \\
University of Nevada, Las Vegas \\
4505 S. Maryland Pkwy \\
Las Vegas, NV, 89154-4002, USA}
\affiliation{Theoretical Division, Los Alamos National Laboratory}

\author[0000-0002-1954-8864]{Sergei Dyda}
\affiliation{Department of Astronomy \\
University of Virginia \\
530 McCormick Rd.\\
Charlottesville, VA 22904, USA}

\author[0000-0003-3616-6822]{Zhaohuan Zhu}
\affiliation{Department of Physics \& Astronomy \\
University of Nevada, Las Vegas \\
4505 S. Maryland Pkwy \\
Las Vegas, NV, 89154-4002, USA}

\begin{abstract}
Local thermal instability can plausibly explain the formation of multiphase gas in many different astrophysical
environments, but the theory of local TI is only well understood in the optically thin limit of the equations 
of radiation hydrodynamics (RHD).  Here, we lay groundwork for transitioning from this limit to a full RHD treatment 
assuming a gray opacity formalism.  We consider a situation where the gas becomes thermally unstable due to 
the hardening of the radiation field when the main radiative processes are free-free cooling and Compton heating.
We identify two ways in which this can happen:
(i) when the Compton temperature increases with time, through a rise in either the intensity or energy 
of a hard X-ray component; and 
(ii) when attenuation reduces the flux of the thermal component so that the Compton temperature increases 
with depth through the slab. Both ways likely occur in the broad line region of active galactic nuclei where columns 
of gas can be ionization bounded. In such instances where attenuation is significant, thermal equilibrium solution curves 
become position-dependent and it no longer suffices to assess the stability of an irradiated column of gas at all depths 
using a single equilibrium curve. We demonstrate how to analyze a new equilibrium curve --- the attenuation curve --- 
for this purpose, and we show that by Field's instability criterion, a negative slope along this curve indicates that constant density slabs are thermally unstable whenever the gas temperature increases with depth.
\end{abstract}
frad
\section{Introduction}
Several outstanding problems in the theory of active galactic nuclei (AGN) are finally becoming tractable due to the rapid progress that has been made in recent years in solving the equations of radiation hydrodynamics \citep[RHD; e.g., ][]{Jiang12,Davis12,Jiang14,Ryan15,Ryan20,Jiang21}.  Prime among them is the structure of the broad line region (BLR), as this is a strongly radiating flow by definition.  
In the class of models popularized by the AGN unification model, only a small volume of the BLR --- the gas within BLR clouds --- contributes to the line emission \citep[for a review, see][]{Netzer15}.
While there is still debate as to the relative contribution of matter bounded clouds (i.e. clouds optically thin to the ionizing continuum so that hydrogen is fully ionized) to the line luminosity \citep[e.g.,][]{Collin-Souffrin88, Shields95, Snedden07},
photoionization modeling generally requires BLR clouds to be ionization bounded (i.e. moderately optically thick) to account for the observed properties of the broad lines \citep[for textbook accounts, see][]{Krolik99, Osterbrock06}.
Testing the viability of BLR cloud models from first principles therefore requires solving the equations of RHD at a minimum.

\cite{Proga14} and \cite{Dyda20} reported on results from a preliminary
investigation of BLR clouds that are ablated by an intense radiation field 
in the gray limit of RHD.
The main focus of these studies was on assessing the role of the type 
and magnitude of the opacity on the evolution of a pre-existing cloud. 
Although these simulations did not include Compton processes, they confirmed the early analytic results of \cite{Krolik88} and \cite{Mathews90}: BLR clouds are highly prone to disruption and various hydrodynamical instabilities. 
\cite{Proga14} noted that optically thin clouds survived the longest.  
Therefore, in subsequent studies, we focused on the optically thin regime by instead solving the equations of non-adiabatic gas dynamics, which allowed including other important radiative processes.  In \cite{PW15}, for example, we studied the formation dynamics of matter bounded BLR clouds in a radiative environment where line cooling and Compton heating resulted in an equilibrium curve (S-curve hereafter) that is thermally unstable.  We demonstrated that in non-isotropic radiation fields, the nonlinear regime of thermal instability \citep[TI;][]{Field65} is associated with cloud acceleration because the line opacity increases as unstable gas condenses \citep[see also][]{WP16}. 

Here, we return to the equations of RHD but now with Compton heating included to allow for the possibility that cloud formation can be triggered due to thermal instability, thus alleviating the need to assume a pre-existing cloud as a part of the initial conditions in RHD calculations.
In \S{2}, we derive a net cooling function that allows us to bridge our previous BLR cloud studies, as it permits studying TI in RHD.  By then focusing our analysis on an irradiated column of gas with constant density, in \S{3} we show how the effects of attenuation modify the basic stability analysis of TI, leading to a new type of equilibrium curve and a physical interpretation of its associated instability criterion.

\section{Theory}
Local TI refers to the exponential growth of linear wave modes in a homogenous plasma.  The entropy mode is sometimes referred to as a trivial mode since it is locally non-propagating, meaning that for a given flow field, it will simply be advected along streamlines.  Only this mode has its stability determined by Balbus's criterion for TI \citep{Balbus86},  
\beq
\left[\pd{(\mathcal{L}/T)}{T}\right]_p < 0,
\label{eq:Balbus_criterion}
\seq
where the subscript notation indicates that the temperature derivative is taken at constant pressure. 
Here, $\mathcal{L}$ is the radiative loss function that enters the first law of thermodynamics, commonly written for gas dynamics as
\beq
   \rho\f{D\mathcal{E}}{Dt} = -p\mathbf{\nabla} \cdot \gv{v} - \rho \mathcal{L} - \mathbf{\nabla} \cdot \gv{q}.
   \label{eq:energy}
\seq
The variables $\rho$, $\gv{v}$, and $p$ are the gas density, velocity,and pressure, respectively, 
$\mathcal{E} = p\rho^{-1}/(\gamma - 1)$ is the gas internal energy(with adiabatic index $\gamma$, taken to be $5/3$ here), 
$\gv{q}$ isthe heat flux due to thermal conduction, and $D/Dt$ is the comoving frame derivative.  We note that
\eqref{eq:Balbus_criterion} will no longer determine the stability of an irradiated gas column at high optical depths, as condensation growth will be suppressed in a radiation diffusion regime; we return to this point in \S{4}.  As argued in Appendix~A, \eqref{eq:Balbus_criterion} will remain valid provided the mean free path for radiation is much greater than both the cooling length of the gas and the gradient length scales of $\rho$, $\gv{v}$, and $p$.  Notice that this instability criterion reduces to Field's criterion, $\left[\pdtext{\mathcal{L}}{T}\right]_p < 0$, for gas in thermal equilibrium with $\mathcal{L} = 0$.

The goal of this section is to arrive at a function $\mathcal{L}$ from first principles, beginning with the source term $S(I,\gv{n})$ describing the interactions between gas and radiation in the radiation transfer equation, 
\beq
\pdif{I}{t} + c\,\gv{n}\cdot\mathbf{\nabla}I = S(I,\gv{n}),
\seq
where $I$ is the frequency-integrated lab frame specific intensity.  In the mixed-frame approach, $S(I,\gv{n})$ is specified in the comoving frame, where it is given by $S(I_0,\gv{n}_0)$.  To leading order, $S(I,\gv{n})$ and $S(I_0,\gv{n}_0)$ differ by $\mathcal{O}(v/c)$ terms, and these terms will be negligible for the dynamics associated with TI whenever the sound speed satisfies $c_s \ll c$.  This is because in the isobaric regime of TI, the condensation velocity is always highly subsonic, while in the nonisobaric regime it is at most $v \sim c_s$ \citep{Waters19a}.
Thus, for the purpose of stability analysis, there is no need to distinguish between these frames.

Assuming local thermodynamic equilibrium (LTE) and using a gray treatment of radiation, the (frequency-integrated) source term is \citep[see e.g.,][]{Jiang21}
\begin{eqnarray}
S(I,\gv{n}) = &&
c\rho\kappa_{aR}\left(\f{a T^4}{4\pi} - I \right)  
+ c\rho\kappa_e(J-I) \nonumber\\
&& +\:c\rho(\kappa_{aP}-\kappa_{aR})\left(\f{a T^4}{4\pi} - J \right) \nonumber \\
&& +\:c\rho\kappa_e\f{4(T-T_{\rm C,eff})}{T_e}J,
\label{eq:S_I}
\end{eqnarray}
where $\kappa_{aR}$ and $\kappa_{aP}$ are the Rosseland and Planck mean absorption opacities, $\kappa_e$ is the electron scattering opacity, and $J$ is the mean intensity defined as, 
\beq
J = \f{1}{4\pi} \int I\,d\Omega.
\label{eq:JfromI}
\seq
The last term in \eqref{eq:S_I} is an approximate way to account for energy exchange through Compton scattering. 
In previous works that use this term \citep[e.g.][]{Hirose09,Jiang13}, $T_{\rm C,eff}$ corresponds to the temperature of the thermal radiation alone, i.e. 
$T_{\rm C,eff} = T_r \equiv (E_r/a)^{1/4}$,
with $E_r$ being the energy density of that radiation, $4\pi J/c$.  As will made clear in \S{3}, the heating due to this term is not efficient enough to lead to TI.  What is needed is a non-thermal component of higher energy radiation.

\subsection{Gray 2-band treatment of Compton heating}
We first define the thermal radiation component as the frequencies extending through the soft X-ray band, 
\beq
J_{\rm s} = \int_0^\nux J_\nu d\nu,
\seq
i.e. up to some frequency $\nux$ that fully covers the blackbody tail of this distribution.
The presence of high energy photons can be accounted for 
in a gray formalism in various simplified ways.  
For example, the mean intensity of hard X-rays can be assumed to be time dependent by taking
$J_{\rm h} = \int_\nux^\infty J_\nu d\nu$ to vary in proportion to $J_{\rm s}$.
Even simpler, the hard X-ray mean intensity can be parameterized in terms of the initial value of $J_{\rm s}$, $J_0 \equiv J_{\rm s}(0) = \int_0^\nux J_\nu \, d\nu \lvert_{t =0}$, allowing the time dependence of $T_{\rm C,eff}$ to be specified using two additional free parameters:
$T_{\rm C,h}$, the Compton temperature corresponding to the hard X-ray component (see \eqref{eq:Tchi}), and $f_{\rm h} \equiv J_{\rm h}/J_0$.
This choice is quite physical and should be a good approximation that applies to the BLR whenever a column of gas is optically thin to this high energy component. 
The radiation source term then becomes 
\begin{eqnarray}
S(I,\gv{n}) = &&
c\rho\kappa_{aR}\left(\f{a T^4}{4\pi} - I_{\rm s} \right)  
+ c\rho\kappa_e(J_{\rm s}-I_{\rm s}) \nonumber\\
&& +\:c\rho(\kappa_{aP}-\kappa_{aR})\left(\f{a T^4}{4\pi} - J_{\rm s} \right) \nonumber \\
&& +\:c\rho\kappa_e\f{4(T-T_{\rm C,eff})}{T_e}(J_{\rm s} + J_{\rm h}),
\label{eq:S_Inew}
\end{eqnarray}
where, as shown in Appendix~B,
\beq
T_{\rm{C,eff}} =\f{T_r\, E_r/E_{r,0} + f_{\rm h} T_{\rm C,h}}{E_r/E_{r,0} + f_{\rm h}}.
\label{eq:Tceff}
\seq
Here, $E_{r,0} = 4\pi J_0/c$ is the initial value of the radiation energy density.

\subsection{Net cooling function in RHD}
By \eqref{eq:JfromI}, we see that the two $\kappa_{aR}$ terms in \eqref{eq:S_Inew} cancel upon integrating over solid angle, $\Omega$.
The source term in the energy equation, 
$S_E \equiv \int S(I,\gv{n}) \,d\Omega$, can therefore be written
\beq
S_E  = \rho \kappa_{aP}(a T^4-E_r) + A_C\rho\,(E_r + f_{\rm h}\,E_{r,0})(T-T_{\rm C,eff}),
\seq
where $E_r$ corresponds only to the thermal radiation,
\beq E_r = \f{4\pi J_{\rm s}}{c},
\seq 
and $A_C = 4\,k\,\kappa_e/(m_e\,c^2)$ is a constant.  Note that by the definition of $f_{\rm h}$, $f_{\rm h} E_{r,0} = 4\pi J_{\rm h}/c$.  With the source term for the momentum equation given by 
\beq
\mathbf{S}_F  = -\frac{\rho(\kappa_e+\kappa_{aR})}{c}\mathbf{F}_r,
\seq
we can now write down the equations of RHD:
\begin{eqnarray}
&&\pdif{\rho}{t} + \mathbf{\nabla} \cdot \left(\rho \gv{v} \right) = 0\label{eq:d_cons}, \\
&&\pdif{\left(\rho\gv{v}\right)}{t} + \mathbf{\nabla} \cdot \left( \rho\gv{v} 
\gv{v} + p I\right) =  - \mathbf{S}_F
\label{eq:m_cons}, \\
&&\pdif{E}{t} + \mathbf{\nabla} \cdot \left[(E + p) \gv{v}\right] = -c S_E - \mathbf{\nabla} \cdot \gv{q}
\label{eq:E_cons}. 
\end{eqnarray}
Comparing \eqref{eq:E_cons} with \eqref{eq:energy}, we have the correspondence $\rho \mathcal{L} = c S_E$.  
Upon adopting a Kramers' opacity appropriate for free-free absorption to use for the Planck mean opacity,
\beq
\kappa_{aP} = A_{\rm ff} \rho\, T^{-3.5},
\label{eq:kappa_ff}
\seq 
we finally arrive at the net cooling function in our RHD framework (noting again that $T_{\rm C,eff}$ is given by \eqref{eq:Tceff}):
\beq
\mathcal{L} = c\,A_{\rm ff} \rho\, T^{-3.5}(aT^4 - E_r) + A_C(E_r + f_{\rm h}\,E_{r,0})(T-T_{\rm C,eff}).
\label{eq:rhoL}
\seq 
In Appendix~A, we justify using \eqref{eq:Balbus_criterion}, which is formally derived under the optically thin approximation, to analyze the stability of this net cooling function. 

\section{Results}
It has been known for over four decades that the thermal state of irradiated gas is determined by the shape of the incident spectrum together with the value of the ionization parameter, $\Xi$ \citep{Krolik81,London81}. 
To the best of our knowledge, all previous dynamical calculations of TI have relied upon the assumption that the shape of the incident spectrum is fixed in both time and space.
We now analyze \eqref{eq:rhoL} to reveal how, by relaxing this assumption,  
an irradiated slab can transition from being thermally stable to unstable and how instability first sets in deep within the slab.

An S-curve can develop an unstable branch whenever the transition from one dominant heating and cooling process to another is too abrupt.  
From \eqref{eq:rhoL}, the ratio of the free-free cooling and Compton heating timescales is 
\beq
\f{t_{\rm cool}}{t_{\rm heat}} = \f{A_C(E_r + f_{\rm h} E_{r,0})T_{\rm C,eff}}{a c A_{\rm ff} \rho \sqrt{T}}
\propto \xi\,  T^{-\f{1}{2}} T_{\rm C,eff},
\label{eq:timescales}
\seq 
where the proportionality relation comes from introducing the density ionization parameter, $\xi$ (defined in \S{3.1} and \eqref{eq:xiEr}). 
At low enough $\xi$ where free-free processes dominate to give $t_{\rm cool} \ll t_{\rm heat}$, the gas can remain stable.  
An unstable branch arises where Compton heating overtakes free-free cooling in \eqref{eq:rhoL}. 
For a thermally stable region within a column of gas, the temperature is fixed by the local value of $\xi$ in that region, so the only way to increase the ratio $t_{\rm cool}/t_{\rm heat}$ is by raising $T_{\rm C,eff}$.  Examining \eqref{eq:Tceff} in the
optically thin limit ($E_r/E_{r,0}=1$) and assuming $T_{r,0}/T_{\rm C,h} \ll f_{\rm h} \lesssim 1$, we find
\beq
T_{\rm{C,eff}} =\f{f_{\rm h}\,T_{\rm C,h}}{1 + f_{\rm h}}.
\label{eq:Tceff_limit1}
\seq
Hence, $T_{\rm C,eff}$ will increase due to either an increase in the temperature ($T_{\rm C,h}$) or mean intensity ($f_{\rm h}$) of high energy photons. 
If the higher value of $T_{\rm C,eff}$ raises $\xi\,T^{-1/2}$ enough such that $t_{\rm cool}/t_{\rm heat}$ continues to increase, the gas has transitioned into a thermally unstable state.

When the optical depth through the slab approaches unity,  
$T_{\rm C,eff}$ will increase without any changes to $T_{\rm C,h}$ or $f_{\rm h}$ due to attenuation.
That this can lead to TI has been noted already by \cite{Goncalves07} in the context of photoionization modeling, as it is purely a radiative transfer effect.
This cause of TI is readily understood from \eqref{eq:Tceff}.
In the limit $f_{\rm h} T_{\rm C,h} \gg T_r E_r/E_{r,0}$ and
neglecting emission through the slab so that  
$E_r/E_{r,0} \approx e^{-\tau}$, where $\tau = \int \kappa \, \rho\, dx$ is the optical depth (with $\kappa = \kappa_e + \kappa_{aP}$), one finds
\beq
T_{\rm{C,eff}} = \f{f_{\rm h} T_{\rm C,h}}{e^{-\tau} + f_{\rm h}}.
\label{eq:Tceff_limit1b}
\seq
We see that \eqref{eq:Tceff_limit1} is recovered for $\tau \ll 1$.  As $\tau$ increases beyond $1$, however, $T_{\rm{C,eff}}$ exponentially approaches $T_{\rm C,h}$, its maximum value.

To summarize, \eqref{eq:timescales} shows that the basic instability mechanism under investigation is an increase in $T_{\rm C,eff}$, and this increase can occur in two different ways:
(i) through temporal changes in the incoming radiation field, i.e. from flux or spectral variability; and
(ii) by a non-temporal route: the spectral hardening of the radiation field with depth.
As demonstrated in \S{3.3}, the parameter space for route (ii) to lead to TI is naturally encountered as an evolutionary phase of route (i).  Put differently, when $T_{\rm C,eff}$ is too close to $T_r$, neither route (i) or (ii) can lead to TI, but as $T_{\rm C,h}$ or $f_{\rm h}$ increase with time, the additional spectral hardening with depth makes route (ii) a possibility before route (i).

\subsection{Position dependent S-curves}
In X-ray astronomy, it is customary to analyze TI in the phase space spanned by the temperature and either the density or pressure ionization parameter, $\xi$ or $\Xi$, respectively.
To express these in terms of $E_r$, first recall that only the soft X-ray flux is treated as thermal radiation, i.e. $E_r \equiv (4\pi/c)J_{\rm s}$.  The total ionizing flux is $F_{\rm{ion}} = F_{\rm s} + F_{\rm h}$,
which for a point source, is related to the total mean intensity $J$ as $F_{\rm{ion}} = 4\,\pi\,J$, or 
$F_{\rm{ion}} = 4\,\pi\,(J_{\rm s} + J_{\rm h}) 
= 4\,\pi\,J_0(E_r/E_{r,0} + f_{\rm h})$.
From the definition
$\xi \equiv 4\pi F_{\rm{ion}}/n$ 
where $n$ the gas number density, we have that $\xi = (4\,\pi)^2 (J_0/n)(E_r/E_{r,0} + f_{\rm h})$.
Taking $n_0$ to be the number density at the face of a slab where $E_r = E_{r,0}$, the incident mean intensity can be parameterized by $\xi_0 = (4\,\pi)^2 (J_0/n_0)(1 + f_{\rm h})$.
Combining the above relations gives, by the definition
$\Xi \equiv (F_{\rm{ion}}/c)/p$, 
\beq
\Xi = \f{1}{4\,\pi\,c\,k} \f{\xi}{T},
\label{eq:bigXi}
\seq
where
\beq
\xi = \f{4\pi\,c}{n}(E_r + f_{\rm h}E_{r,0}),
\label{eq:xiEr}
\seq
and
\beq
E_{r,0} = \f{\xi_0\,n_0}{4\pi\,c\,(1+f_{\rm h})}.
\label{eq:Er0}
\seq
From \eqref{eq:xiEr}, we see that for a gas column exposed to a given mean intensity, a rise in density is not the only way the ionization can decrease; as noted above,
attenuation will also reduce $\xi$ by resulting in $E_r < E_{r,0}$.  
This represents a major departure from the textbook approach to analyzing TI, where a single S-curve 
suffices to delineate the unstable parameter space. 
That S-curves become position dependent in attenuated gas was also pointed out by \cite{Goncalves07}.

Conveniently, there is an analytic expression for these  S-curves. 
Setting $\mathcal{L}=0$ in \eqref{eq:rhoL}, solving for $\rho$, and equating this with $\rho = \bar{m} (k\,T\,\Xi)^{-1} (E_r + f_{\rm h} E_{r,0})$ from \eqref{eq:bigXi} gives,
\beq
\Xi_S(T) = \f{\bar{m}\,c\,A_{\rm ff}}{A_C} \f{a\,T^4 - E_r}{k(T_{\rm C,eff}-T)} \,T^{-4.5}.
\label{eq:Scurve_xi}
\seq
Because $T_{\rm C,eff} = T_{\rm C,eff}(E_r)$, 
the position dependence is solely due the local value of $E_r = E_r(\gv{x})$. 
For gas columns with $\tau \ll 1$, a negligible amount of attenuation implies $E_r = E_{r,0}$, so S-curves in the optically thin limit have $E_r$ given by \eqref{eq:Er0} and $\xi = \xi_0 n_0/n$ by \eqref{eq:xiEr}.
The S-curve is so named because on the $(T,\Xi)$-plane, it has a negative-sloping unstable branch connecting low and high temperature positive-sloping stable branches.
At small and large $\Xi$ these stable branches correspond to the asymptotic limits obvious from \eqref{eq:Scurve_xi}: as $\Xi_S \rightarrow \infty$, $T \rightarrow T_{\rm{C,eff}}$, while as $\Xi_S \rightarrow 0$, $T \rightarrow T_r \equiv (E_r/a)^{1/4}$.

\begin{figure*}
  \centering
  \includegraphics[width=0.98\textwidth]{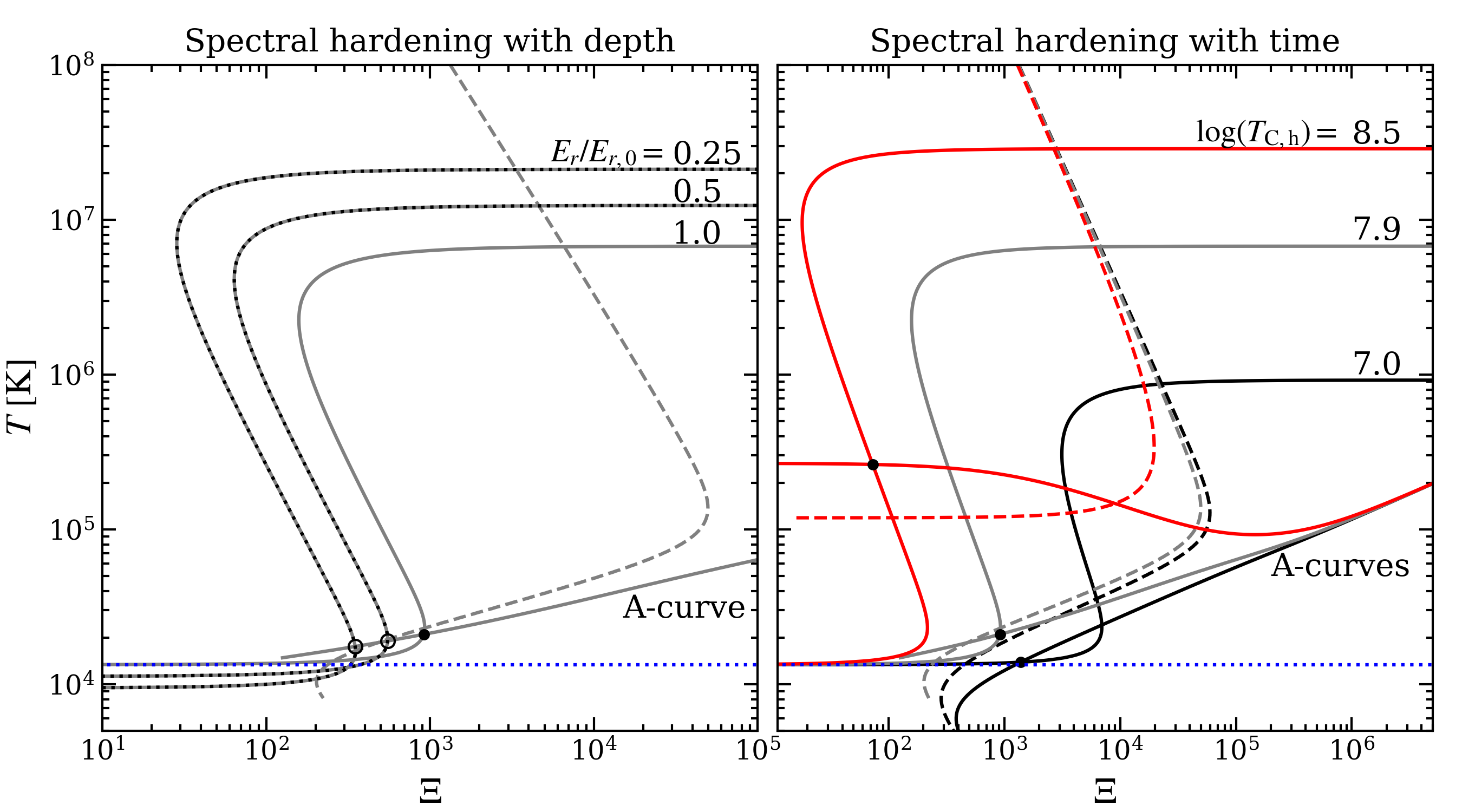}
  \caption{Examination of the two ways Compton heating can lead to TI in RHD.  Either A-curves, along with their corresponding Balbus contours (dashed lines), or S-curves can be used to assess stability.  Balbus contours for S-curves are not shown since negative slopes indicate instability.
  Horizontal dotted lines denote the initial radiation temperature, $T_{\rm{r,0}}$, of the thermal soft X-ray flux.
  \emph{Right panel}: Route (i) to TI: $T_{\rm C, eff}$ is increased with time by raising $T_{\rm C,h}$ for a fixed $f_{\rm h}$ (set to 0.1).  The dot on the black S-curve marks the irradiated face of an initially stable slab.  It will move to the black dot on the gray (red) S-curve if $\log T_{\rm C,h}$ is increased to 7.9 (8.5).  In the $\log T_{\rm C,h} = 7.9$ (8.5) case, the face of the slab remains stable (becomes unstable) because the S-curve has a positive (negative) slope at that point; equivalently, it lies beneath (above) the Balbus contour for that A-curve.
  \emph{Left panel}: Route (ii) to TI: $T_{\rm C, eff}$ increases with depth as a consequence of spectral hardening due to attenuation. 
  The black dot marks the face of the slab corresponding to the $\log T_{\rm C,h} = 7.9$ S-curve and A-curve from the right panel.  The two open circles mark the $(T,\Xi)$ values at the depths where the smaller values of $E_r/E_{r,0}$ are reached.  Thermal stability can be assessed using a single A-curve: a transition to instability occurs between the two open circles once the Balbus contour crosses below the A-curve.  Alternatively, multiple S-curves can be computed to see this transition.  They are labeled by the value of $E_r/E_{r,0}$ and explicitly show the spectral hardening with depth.  The $E_r/E_{r,0} = 0.25$ point is seen to occupy the unstable branch.
  }
  \label{fig:XiScurves}
\end{figure*}

\subsection{Attenuation curves}
When TI is analyzed using S-curves, smaller values of $\xi = 4\pi F_{\rm ion}/n$ are associated with either smaller ionizing fluxes or higher densities, which is to say that $n\,\xi$ is treated as a proxy for $F_{\rm ion} \propto J$. 
Degeneracy will arise when attenuation becomes important because $J$ is approximately $J_0 e^{-\tau}$, i.e. $n\,\xi \propto J_0 e^{-\tau}$ becomes a proxy for both $J_0$ and the density.  To break the degeneracy, it becomes necessary to supplement the S-curve analysis by focusing on gas at some fixed density.  Specifically, taking $\rho=constant$ in \eqref{eq:xiEr} leads to an alternative stability analysis of an attenuated column of gas, with smaller $\xi$ values corresponding to greater depths through the column. 

A second equilibrium curve, unique to RHD, that we term the \textit{attenuation curve}, or A-curve for short, applies to all gas at a given density.  It corresponds to the $\mathcal{L} = 0$ contour of \eqref{eq:rhoL} after replacing $E_r$ with $\xi$ using \eqref{eq:xiEr}.\footnote{There is no analytic expression for A-curves due to the nonlinear $E_r$-dependence of $T_{\rm C,eff}$ in \eqref{eq:Tceff}.}
In contrast with S-curves, A-curves represent the phase space the column will occupy as the flux drops with depth due to significant absorption or scattering.  
An A-curve is shown in the left panel of Fig.~\ref{fig:XiScurves} for a constant density slab, where it is compared with three S-curves, one computed at the face of the slab (solid line) and two at significant depths into the slab (solid lines with dots overplotted), computed by taking $E_r = 0.5 E_{r,0}$ and $E_r = 0.25 E_{r,0}$ in \eqref{eq:Scurve_xi}.  

We emphasize again that the $\Xi$-axis for an S-curve is a proxy for pressure, while the $\Xi$-axis for an A-curve is a proxy for $E_r$, i.e. these axes are separately built by taking $E_r=constant$ and $n=constant$ in \eqref{eq:xiEr}, respectively.
With $\Xi_0$ specified at the face of the slab (marked by the solid black dot), points along the A-curve with $\Xi < \Xi_0$ correspond to greater depths into the slab.  
The S-curves illustrate the spectral hardening with depth because as $E_r/E_{r,0}$ decreases, the Compton branch of each one tends to higher values of $T_{\rm{C,eff}}$. 
The A-curve reveals that the temperature tends to decrease with greater depth through the column, the behavior expected for an attenuated slab.  

We can now state a key result of this study: an irradiated constant density slab is thermally unstable if, due to attenuation, the temperature \textit{increases} with depth.  This result follows from the mathematical identity
$(\pdtext{T}{p})_\mathcal{L} (\pdtext{p}{\mathcal{L}})_T (\pdtext{\mathcal{L}}{T})_p  = -1$.
As shown in Appendix~C, in the optically thin limit the following version of this identity applies to both S-curves and A-curves:
\beq
\left(\pdif{\log T}{\log \Xi}\right)_\mathcal{L} 
= \f{p}{T}\f{(\pdtext{\mathcal{L}}{p})_T}{(\pdtext{\mathcal{L}}{T})_p}.
\label{eq:Acurve_identity}
\seq 
The left hand side derivative is taken along surfaces of constant $\mathcal{L}$; this equation therefore relates Field's instability criterion, $(\pdtext{\mathcal{L}}{T})_p < 0$, to the slopes of S-curves and A-curves (values of $\pdtext{\log T}{\log \Xi}$ along the contours $\mathcal{L}=0$).
This correspondence between instability according to Field's criterion and
the slope having $(\pdtext{\log T}{\log \Xi})_\mathcal{L} < 0$ holds so long as
$(\pdtext{\mathcal{L}}{p})_T > 0$, which is readily seen to be the case for S-curves but not for A-curves (see Appendix~C).  In other words, in the case of A-curves, it is possible for a slab to be thermally unstable when $(\pdtext{\log T}{\log \Xi})_\mathcal{L} > 0$ because $(\pdtext{\mathcal{L}}{p})_T$ can become negative.
We will show an example of this below.
Focusing on the case $(\pdtext{\mathcal{L}}{p})_T > 0$ shows that unlike for an S-curve, a negative slope for an A-curve physically corresponds to a slab with a temperature that increases with depth.

\subsection{Stability analysis using S-curves and A-curves}
For a stratified or otherwise inhomogeneous column of gas, S-curves are position dependent unless attenuation within the column is negligible, and A-curves are also position dependent unless the column has constant density.  Hence, by focusing on constant density slabs, a stability analysis requiring multiple S-curves is equivalent to one using a single A-curve. Returning to the left panel of Fig.~\ref{fig:XiScurves},
notice that at some depth, corresponding to $E_r/E_{r,0} \approx 0.5$, gas occupies the last stable point on the cold branch of the S-curve (see the open circle left of the black dot).  At greater depths, it occupies the unstable branch.  This conclusion, i.e. that attenuation causes the slab to become unstable at some distance within, can alternatively be arrived at by analyzing the A-curve: a transition to instability occurs when this curve passes above its corresponding Balbus contour (see Appendix~D), shown here as the dashed line.

To demonstrate what is likely a physical scenario that can be studied numerically, consider an initially thermally stable slab that begins to evolve as $T_{\rm{C,eff}}$ increases with time.  At $t=0$, the face of the slab will occupy a single point on the S-curve.  We show such a point as the black dot on the $\log(T_{\rm C,h}) = 7$ S-curve in the right panel of Fig.~\ref{fig:XiScurves}.
In this particular example, we increase $T_{\rm{C,eff}}$ with time by raising $T_{\rm C,h}$ with $f_{\rm h}$ held fixed at 0.1.  The density of the slab will remain constant, at least initially, while the temperature (and hence pressure) will increase until the slab reaches the new equilibrium temperature corresponding to $T_{\rm{C,eff}}(t)$.  We note that $\Xi$ will decrease because $\Xi \propto p^{-1}$.

We consider two cases: the final value of the free parameter $\log(T_{\rm C,h}) $ is (i) 7.9 and (ii) 8.5.  The slab settles on the gray S-curve in case (i) and the red S-curve in case (ii), at the locations marked with a black dot.  In case (ii), the entire slab is now unstable: its face occupies the unstable branch of the S-curve, and the A-curve shows that the attenuated layers at lower $\Xi$ have temperatures slightly greater than at $\Xi_0$, thus satisfying the negative slope criterion for instability (see \eqref{eq:Acurve_identity}).
The analysis of case (i) was already presented when discussing the left panel. 

Notice that for a slab to rise to the temperature of case (ii), it must first pass through the parameter space of case (i).  Whether or not the unstable portion of the slab can undergo TI before the full slab becomes unstable (upon occupying the red S-curve) depends on the timescales involved during this period of heating.  
Explicit slab structure calculations are needed to determine if there can be multiple dynamical outcomes.  

\section{Discussion and Conclusions}
Numerous attempts made over the decades to explain the presence of broad emission lines in AGN spectra are posited on TI being the process that forms the BLR clouds within which many of the lines originate \citep[e.g.,][]{Beltrametti81,Perry85,Shlosman85,Krolik88,Mathews90,Goncalves93,Wang12,Elvis17}.  
Perhaps the most promising cloud-based BLR models have been developed in the framework of magnetohydrodynamics \citep[MHD;][]{EBS92,dKB95,BS17}; of these, only the model by \cite{BS17} is focused on TI, which is envisioned to occur in the atmosphere of a magnetically elevated disk.
Testing the viability of this model
will require solving the equations of radiation MHD (RMHD).
Advances in understanding of the dynamics of TI in MHD have been led by the solar physics community in recent years
\citep[see reviews by][]{Soler22,Antolin22}, especially in the context of solar prominence formation and the coronal rain phenomenon \citep[e.g.,][]{Claes19,Claes20,Hermans21}.  Because these studies have all assumed an optically thin formalism where attenuation is neglected, we stress that the RHD framework and analysis methods presented here are equally applicable to the equations of RMHD.
 
We also emphasize that TI is a process inherent to \textit{somewhat} optically thin media: the growth of condensations will be suppressed in an optically thick, radiation diffusion regime just as condensation growth is suppressed in optically thin media for wavelengths less than the Field length.  In the latter case, the cooling of an unstable entropy mode cannot take place when the heat flux due to thermal conduction is strong enough to maintain a flow of heat from hot to cold across the entire perturbation; this circumstance is what the Field length identifies \citep{Begelman90}.  In a radiation diffusion regime, TI will also be suppressed because radiation will establish LTE and hence wipe out any temperature inhomogeneities.  

The theory developed in this paper extends the optically thin qualification to encompass situations where attenuation is too important to completely ignore.
We retain all the approximations made by \citet{Balbus86}:
the relevant wavenumbers are large such that gas can be treated as homogeneous over length scales that are much smaller than those on which the background flow varies.
In such a regime, the familiar criteria for TI still apply but must be evaluated layer by layer through an attenuated slab.  In lieu of this position-dependent analysis using S-curves,  it is possible to use a single A-curve to assess the stability of a constant density column of gas.  As with S-curves, the slope of A-curves is mathematically related Field's instability criterion.  However, there is no longer a one-to-one correspondence between a negative slope on the $(T,\Xi)$-plane and instability for A-curves.  Instability is less restrictive: an A-curve with a negative slope is unstable to TI but the threshold is actually at some small positive slope, as shown in Fig.~\ref{fig:XiScurves} and discussed in Appendix~C.
Remarkably, this analysis reveals that a negative slope on an A-curve has the physical implication that a constant density slab with an increasing temperature profile is thermally unstable. 

We furthermore demonstrated, again in the case of a constant density slab, how S-curves and A-curves are complimentary in understanding how the slab will evolve under temporal changes in the hardness of the radiation field.  This type of analysis may prove especially useful in understanding the dynamics of more complicated gas distributions.  If, for example, we envision a Compton thick, multiphase slab, all the gas with a given density will occupy a single A-curve, while all the gas at a given mean intensity will occupy a single S-curve.  In subsequent papers of this series, we will apply this analysis to slab structure calculations to understand the nonlinear outcome of TI in RHD.

\acknowledgments 
We are indebted to Yan-Fei Jiang for pointing out that the discrepancy between $4\,k\,T_{\rm C,s}$ and $\E{h\nu}_{B}$ in \eqref{eq:hnu_B} is due to the neglect of
stimulated scattering.  Support for this work was provided by the National Aeronautics and Space Administration under ATP grant NNX14AK44G and TCAN grant 80NSSC21K0496.

\appendix
\section{Applicability of established linear theory}
\eqref{eq:rhoL} has a functional dependence $\mathcal{L} = \mathcal{L}(\rho, T, E_r)$, whereas the stability criteria for local TI were derived assuming $\mathcal{L} = \mathcal{L}(\rho, T)$ under the optically thin approximation.  While we aim to reach an RHD regime in the nonlinear phase of TI where the gas within BLR clouds is optically thick and radiating strongly, here we are only concerned with the background flow out of which the clouds form.  This gas is seen to be optically thin locally, meaning that the wavelengths of unstable entropy modes are much smaller than the mean free path of the X-ray photons. Specifically, adopting $A_{\rm ff} = 5.19\times 10^{24}\,\rm{cm^5\,g^{-2}\,K^{3.5}}$ for the bremsstrahlung constant in \eqref{eq:kappa_ff}, the opacity is dominated by electron scattering at the densities and temperatures considered:
\beq
\f{\kappa_e}{\kappa_{aP}} = 1.2\times 10^5\,n_{11}^{-1} \,T_{5}^{3.5}.
\seq
Here, $n_{11} = n/10^{11} \, {\rm cm}^{-3}$ and $T_5 = T/10^5 \, {\rm K}$ are characteristic values for the BLR.
In examining the stability of an irradiated column of gas, 
$\mathcal{L}(\rho, T, E_r)$ will be reduced back to $\mathcal{L} = \mathcal{L}(\rho, T)$ if we can treat $E_r$ as constant over portions of the column $\Delta x$ satisfying $\Delta x \ll \lambda_{\rm mfp}$, with $ \lambda_{\rm mfp} = (\kappa_e \rho)^{-1} = 5.0 \times 10^{13}\,n_{11}^{-1}\, {\rm cm}$ the mean free path of the soft X-ray photons.
Applying local TI theory also requires the gas to be homogeneous over scales where the growth rate of TI peaks, namely for $\Delta x \gtrsim 10\, \lambda_{\rm cool}$, where $\lambda_{\rm cool} = c_s t_{\rm cool}$ is the cooling length. 
Combining the above requirements, the local approximation holds when $\lambda_{\rm cool} \ll \lambda_{\rm mfp}$.
With $t_{\rm cool} = c_{\rm v} T/(c\,a\,A_{\rm ff}\rho T^{1/2})$ for free-free cooling, we obtain $\lambda_{\rm cool} = 2.6\times 10^9\, T_5\,n_{11}^{-1}\,{\rm cm}$; thus, $\lambda_{\rm cool} \ll \lambda_{\rm mfp}$ is satisfied independent of density.
We further note that the radiation force will not alter the linear theory since the flow will be uniformly accelerated when $\kappa_e \gg \kappa_{aP}$, thereby permitting an analysis in the comoving frame. 

The above length scale hierarchy validates the use of the so-called eikonal approximation (or WKBJ theory) that was used by \citet{Balbus86} and permits applying \eqref{eq:Balbus_criterion} to a time-dependent background flow such as the one considered here (see \S{3.3} and the right panel of Fig.~\ref{fig:XiScurves}).

\section{Parameterization of the mean photon energy}
In our gray treatment of the X-ray irradiation, the time-dependent mean photon energy defines an effective Compton temperature $T_{\rm C,eff}$ through
\beq
\begin{split}
\E{h\nu} \equiv 4\,k\,T_{\rm C,eff} & =\f{\int_0^\infty h\nu J_\nu \, d\nu}{\int_0^\infty J_\nu \, d\nu} \\
          &= \f{\int_0^\nux h\nu J_\nu \, d\nu + \int_\nux^\infty h\nu J_\nu \, d\nu}{J_{\rm s} + J_{\rm h}},
\label{eq:ep0a}
\end{split}
\seq
Here, $\nux$ divides the incident spectrum into two components:
(i) $J_{\rm s} \equiv \int_0^\nux J_\nu \, d\nu$, the thermal component spanning the far UV and soft X-ray energy range for which we solve the radiation transfer equation; and
(ii) $J_{\rm h} \equiv \int_\nux^\infty J_\nu \, d\nu$, the high energy tail (i.e. keV range) that is assumed constant and unevolving.
Both components have an associated Compton temperature, as the terms in the numerator are related to mean photon energies, viz.
\beq
4\,k\,T_{\rm C,s} = \E{h\nu}_{\rm s} = J_{\rm s}^{-1}\int_0^\nux h\nu J_\nu \, d\nu;
\label{eq:TClo}
\seq
\beq
4\,k\,T_{\rm C,h} = \E{h\nu}_{\rm h} =  J_{\rm h}^{-1} \int_\nux^\infty h\nu J_\nu \, d\nu.
\label{eq:Tchi}
\seq
Below we explain how to modify this two-band formalism to make $T_{\rm C,s}$ exactly equal to $T_r = (E_r/a)^{1/4}$, the radiation temperature that is evolved using the equations of RHD.  Meanwhile, $T_{\rm C,h}$ is taken as a free parameter and controls the thermal stability of the gas. 

To express $T_{\rm C,eff}$ in terms of $T_{\rm C,s}$, $T_{\rm C,h}$, and $E_r$, 
we first note that at $t=0$, the second line in \eqref{eq:ep0a} can be written as
\beq
T_{\rm C,eff}(t=0) = \f{T_{\rm C,s} + f_{\rm h} T_{\rm C,h}}{1 + f_{\rm h}},
\label{eq:ep0b}
\seq
where $f_{\rm h} = J_{\rm h}/J_0$ parameterizes component (ii) in terms of the initial value of component (i), $J_0 \equiv J_{\rm s}(t =0)$.  
Further recognizing that $J_{\rm h} = (E_r/E_{r,0})J_0$, 
with $E_{r,0} = (4\pi/c) J_0$, the time-dependent decomposition then becomes
\beq
T_{\rm C,eff} = \f{T_{\rm C,s} E_r + f_{\rm h} T_{\rm C,h} E_{r,0}}{E_r + f_{\rm h} E_{r,0}}.
\label{eq:Tceff0}
\seq
This reduces to $T_{\rm C,s}$ for $f_{\rm h}=0$, and thus corresponds to the simplified treatment of Compton scattering that has been used in the past \citep[e.g.][]{Hirose09,Jiang13} when $T_{\rm C,s} = T_r$.  However, the equality $T_{\rm C,s} = T_r$ is not obtained using the definition of $\E{h\nu}$ in \eqref{eq:ep0a}. The mean photon energy for a Planck spectrum, $J_\nu = B_\nu$, is
\beq
\E{h\nu}_{B} \equiv \f{\int_0^\infty h\nu B_\nu \, d\nu}{\int_0^\infty B_\nu \, d\nu} = \f{360\,\zeta(5)}{\pi^4} k\,T_r \approx 3.823\, k\,T_r,
\label{eq:hnu_B}
\seq
where $\zeta(s)$ is the Riemann zeta function.  With this definition, net heat exchange will take place between gas and radiation at temperature $T_r$.  When Compton scattering is treated using the Kompaneets equation (the frequency diffusion limit), the mean photon energy is instead given by \citep[see][]{Blaes01}
\beq 
\E{h\nu} =\f{\int_0^\infty h\nu J_\nu[1 + (c^2/2h) J_\nu/\nu^3] \, d\nu}{\int_0^\infty J_\nu \, d\nu} ,
\seq
with the factor $(c^2/2h) J_\nu/\nu^3$ accounting for stimulated scattering.  With this definition in place of \eqref{eq:ep0a}, we still arrive at the parametrization given in \eqref{eq:Tceff0} but now with $T_{\rm C,s} = T_r$, which brings us to \eqref{eq:Tceff}.

\section{The slopes of equilibrium curves in RHD}
Beginning with the identity quoted in \S{3.2},
$(\pdtext{T}{p})_\mathcal{L} (\pdtext{p}{\mathcal{L}})_T (\pdtext{\mathcal{L}}{T})_p  = -1$,
we use the chain rule on the first derivative to introduce the slope along equilibrium curves, $(\pdtext{\log T}{\log \Xi})_\mathcal{L}$:
\beq
\left(\pd{T}{p}\right)_\mathcal{L} = \f{T}{p} \left(\pd{\log T}{\log \Xi}\right)_\mathcal{L} \left(\pd{\log\Xi}{\log p}\right)_\mathcal{L}.
\seq
By \eqref{eq:xiEr} and the definition $\Xi = (4\,\pi\,c\,k)^{-1} \xi/T$, we have $(\pdtext{\log\Xi}{\log p})_\mathcal{L} =  \Xi^{-1}(\pdtext{E_r}{p})_\mathcal{L} - 1$.  Meanwhile, for the derivative $(\pdtext{p}{\mathcal{L}})_T$, the chain rule gives
\beq 
\left(\pd{\mathcal{L}}{p}\right)_T = 
\left(\pd{\mathcal{L}}{\rho}\pd{\rho}{p}\right)_T + \left(\pd{\mathcal{L}}{E_r}\pd{E_r}{p} \right)_T.
\seq
By the ideal gas law, $(\pdtext{\rho}{p})_T = \rho/p$.  With these substitutions, the original identity becomes 
\beq
\left[1 - \f{1}{\Xi}\left(\pd{E_r}{p}\right)_\mathcal{L}\right]
\left(\pd{\log T}{\log \Xi}\right)_\mathcal{L} 
= \f{\rho(\pdtext{\mathcal{L}}{\rho})_T + p(\pdtext{\mathcal{L}}{E_r})_T (\pdtext{E_r}{p})_T}{T(\pdtext{\mathcal{L}}{T})_p}.
\label{eq:identityA}
\seq 
This equation gives the relation between Field's instability criterion, $(\pdtext{\mathcal{L}}{T})_p < 0$, and the slopes of S-curves and A-curves.  
In the case of S-curves, $E_r$ is independent of $p$ and this equation reduces to 
\beq
\left(\pd{\log T}{\log \Xi}\right)_\mathcal{L} 
= \f{\rho(\pdtext{\mathcal{L}}{\rho})_T}{T(\pdtext{\mathcal{L}}{T})_p}.
\label{eq:identityS}
\seq 
The derivative $(\pdtext{\mathcal{L}}{\rho})_T$ is readily seen to be positive by \eqref{eq:rhoL} and this is generically true for astrophysical cooling functions (\citealt{Balbus89}; see also Appendix~B of \citealt{Waters19a}), giving a one-to-one relation between Field's criterion and the sign of the slope of S-curves.  For A-curves, the bracketed term on the left hand side of \eqref{eq:identityA} can be shown to be positive but the derivative $(\pdtext{\mathcal{L}}{E_r})_T$ can be negative and change the sign of the numerator on the right hand side.
The one-to-one correspondence is therefore lost.  However,
in the optically thin limit, $\pdtext{E_r}{p}$ approaches 0, showing that \eqref{eq:identityS} holds for A-curves in this limit, a fact relied upon in \S{3.2}. 

\section{Balbus Contours}
Diagnosing instability using \eqref{eq:Balbus_criterion} entails computing the location in a slab where the quantity 
$[\pdtext{(\mathcal{L}/T)}{T}]_p$ changes sign, which can be done by plotting the Balbus contour, defined as locations in phase space where $[\pdtext{(\mathcal{L}/T)}{T}]_p = 0$.  Dividing \eqref{eq:rhoL} by $T$ and replacing $\rho$ with $p$ using the ideal gas law (so that $p$ can be held constant while taking the derivative with respect to $T$) gives
\beq\left[\pd{(\mathcal{L}/T)}{T}\right]_p  =  -\f{3}{2} c\,A_{\rm ff} \rho\, T^{-5.5}\left(aT^4 - \f{11}{3} E_r\right)
 + A_C\, T^{-2}\, (E_r + f_{\rm h}\,E_{r,0})\,T_{\rm C,eff}.
\label{eq:Bcontour}
\seq
 Both S-curves and A-curves have corresponding Balbus contours.  The analytic expression that accompanies \eqref{eq:Scurve_xi} is found by setting  $[\partial(\mathcal{L}/T)/\partial T]_p=0$ in \eqref{eq:Bcontour}, solving for $\rho$ and, as before, setting this expression equal to $\rho = \bar{m} (k\,T\,\Xi)^{-1} (E_r + f_{\rm h} E_{r,0})$ from \eqref{eq:bigXi} to give
\beq
\Xi_B(T) = \f{3}{2}\f{\bar{m}\,c\,A_{\rm ff}}{A_C} \f{a\,T^4 - (11/3) E_r}{k\,T_{\rm C,eff}} \,T^{-4.5}.
\label{eq:Bcurve_Xi}
\seq
By instead using \eqref{eq:bigXi} to eliminate $E_r$ from \eqref{eq:Bcontour}, the equation that results from setting $[\partial(\mathcal{L}/T)/\partial T]_p=0$ yields the Balbus contours corresponding to A-curves that are plotted in Fig.~\ref{fig:XiScurves}.
The utility of using Balbus contours to identify where gas can be thermally unstable is that they also reveal the parameter space for TI when gas departs from the equilibrium curve (the contour $\mathcal{L} = 0$).

\end{document}